\documentclass[useAMS,usenatbib]{mn2e}
\usepackage{aas_macros}
\usepackage{graphicx}
\usepackage{amsmath}
\usepackage{amsfonts}
\usepackage{amssymb}%
\begin{document}

\title[{Comptonization of photons in relativistic outflows}]{Comptonization of photons near the photosphere of relativistic outflows}
\author[A.G. Aksenov, R. Ruffini and G. V. Vereshchagin]{A.G. Aksenov$^{1}$\thanks{E-mail:
aksenov@icad.org.ru (AGA); ruffini@icra.it (RR); veresh@icra.it (GVV)}, R. Ruffini$^{2,3}$ and G. V. Vereshchagin$^{2,3}$ \\
$^{1}$Institute for Computer-Aided Design, Russian Academy of Sciences, Vtoraya Brestskaya 19/18, Moscow, 123056, Russia \\
$^{2}$ICRANet, 65122, p.le della Repubblica, 10, Pescara Italy \\
$^{3}$University of Roma ``Sapienza'', 00185, p.le A. Moro 5, Rome, Italy}



\maketitle

\label{firstpage}

\begin{abstract}
We consider the formation of photon spectrum at the photosphere of
ultrarelativistically expanding outflow. We use the Fokker-Planck
approximation to the Boltzmann equation, and obtain the generalized Kompaneets
equation which takes into account anisotropic distribution of photons
developed near the photosphere. This equation is solved numerically for relativistic steady wind and the observed spectrum is found in agreement with previous studies. We also study the photospheric emission for different temperature dependences on radius in such outflows. In particular, we found that for $T\propto r^{-2}$ the Band low energy photon index of the observed spectrum is $\alpha\simeq -1$, as typically observed in Gamma Ray Bursts.
\end{abstract}


\section{Introduction}

Gamma Ray Burst (GRB) emission originates from plasma which expands
relativistically from initial optically thick phase. It necessarily contains
the photospheric component which appears when plasma becomes transparent to
radiation initially trapped in it
\citep{1986ApJ...308L..47G, 1986ApJ...308L..43P, 1990ApJ...363..218P,
1990ApJ...365L..55S}.

The fireshell model, see. e.g. \citet{2009AIPC.1132..199R} pays
special attention to this photospheric component. It is identified as P-GRB
(proper GRB). Both its energetics and time separation from the peak of the
afterglow are predicted, and comparison with observations allowed
identification of this component in many GRBs (970228, 991216, 031203, 050315,
050509B, 060218, 060607A, 060614, 071227, 090618, 090902B, 090423 and others).

The impossibility of explaining observed hard low energy slopes of spectra with traditional
synchrotron shock models revived the interest in the photospheric emission
from relativistic outflows in the recent literature
\citep{2000ApJ...530..292M, 2002ApJ...578..812M,
2002MNRAS.336.1271D, 2007ApJ...664L...1P, 2010MNRAS.407.1033B,
2011ApJ...737...68B, 2011MNRAS.415.3693R, 2011ApJ...732...49P}. Reprocessing
of thermal emission assumed to originate from the photosphere by nonthermal
distribution of electrons is in the basis of more complicated models with
dissipation near the photosphere \citep{2005ApJ...628..847R,
2006A&A...457..763G, 2011MNRAS.415.1663T, 2011MNRAS.415.3693R,
2013ApJ...764..143V, 2012ApJ...756..174L}. Such models successfully reproduce
both high energy and low energy parts of observed spectra in some bursts: e.g.
GRB090902B, see \cite{2011MNRAS.415.3693R}. However, the physics of
dissipation is poorly understood. Fine tuning is involved, since such
dissipation is required to occur very near the photosphere, see e.g.
\cite{2013ApJ...764..143V}.

Several papers claim deviations of theoretically computed spectrum, coming
from nondissipative photospheres of relativistic outflows, from the Planck
shape at low energies. \cite{2010MNRAS.407.1033B} found the low energy photon
index $\alpha=0.4$\ (in contrast with $\alpha=1$ for Planck spectrum) solving
the radiative transfer equations by Monte Carlo method in the steady
relativistic wind.
\cite{2008ApJ...682..463P} and \cite{2011ApJ...732...49P} found both
analytically and numerically flattening of the spectrum, considering late time
emission from the relativistic wind when it switches off with $\alpha
\rightarrow0$ for late emission. This result is obtained considering the
probability density function of the last scattering.
\cite{2013ApJ...772...11R} found that the increased power at low energies with
respect to the Planck spectrum (photon indices are, respectively $\alpha=0.82$ and $0.34$
for accelerating and coasting photon thick instantaneous spectra) is the
result of temperature distribution across the Photospheric EQuiTemporal
Surface (EQTS) \citep{2001A&A...368..377B, 2001ApJ...555L.113R}.

The formation of photon spectrum at the photosphere of ultrarelativistically
expanding outflow can be explained as a combination of several effects.
Firstly, there is a contribution from different angles with respect to the
line of sight of photons, arriving to the observer at the same time from the
EQTS. This effect is purely geometrical. Secondly, photons have different
comoving temperatures at each point of EQTS and the observed spectrum is a
superposition of that comoving spectra, blueshifted to the observer's
reference frame (a multicolor blackbody) \citep{2011ApJ...732...49P}. Thirdly,
local spectral distortions arise due to the decoupling of photons from the plasma
near the photosphere. In this Letter we are concerned with the latter effect.
We use the radiative transfer equation for the steady wind model of
\cite{2011ApJ...737...68B} with collision integrals obtained as the
Fokker-Planck approximation to exact collision integrals of Compton
scattering. This equation reduces to the classic Kompaneets equation for the
medium at rest. However, here we consider relativistically expanding outflow,
implying that anisotropy of the photon field has to be accounted for in the
collision integral. We solve numerically this equation and find the observed
spectrum of photospheric emission from relativistic wind.

The paper is organized as follows. In Section 2 we present the radiative transfer equation for coasting relativistic wind. In Section 3 we report our numerical results. Conclusions follow.

\section{The radiative transfer equation}

We start with the radiative transfer equation in spherically symmetric steady
(time independent) medium written in comoving frame, with the only exceptions
for radius $r$\ and Lorentz factor $\Gamma$\ being the quantities measured in
the laboratory frame (c.f. \cite{2011ApJ...737...68B}):
\begin{eqnarray}
\frac{\mu+\beta}{1+\beta\mu}\frac{\partial n(r,\epsilon,\mu)}{\partial r}+\frac{1-\mu^{2}}{r}%
\frac{\partial n}{\partial\mu}-\frac{1-\mu^2}{r(1+\beta\mu)}\beta\epsilon\frac{\partial
n}{\partial\epsilon} \nonumber \\
=\frac{1}{\Gamma(1+\beta\mu)}\frac{\eta-\chi n}{1+\beta\mu},
\label{RadTransf}%
\end{eqnarray}
where $n$ is photon occupation number, $\mu=\cos\vartheta$ parametrizes the
angle between the momentum of the photon and the radius vector, $\eta$\ and
$\chi$\ are emission and absorption coefficients, $\epsilon$\ is photon energy.

The collision integral takes into account that photons are scattered by the
moving medium (see \cite{1985PASJ...37..383F, 1997ApJ...488..881P}) \ and
consequently their distribution is anisotropic in the comoving frame:%
\begin{align}
\eta-\chi n &  =\frac{3n_{e}\sigma_{\mathrm{T}}\epsilon}{16\pi}\Bigg\{\left(
1-\frac{2\epsilon}{m_{e}c^{2}}\right)  \label{Coll}\\
&  \times\int do^{\prime}\left[  \left(  1+x^{2}\right)  n^{\prime}-n\right]
\nonumber\\
&  +\frac{2\epsilon}{m_{e}c^{2}}\int do^{\prime}\left(  x^{3}+x\right)
n^{\prime}\nonumber\\
&  +\frac{2kT_{e}}{m_{e}c^{2}}\int do^{\prime}\left(  2x^{3}-3x^{2}%
-2x+1\right)  n^{\prime}\nonumber\\
&  +\frac{1}{m_{e}c^{2}}\int do^{\prime}\left(  1+x^{2}\right)  \left(
1-x\right)  \nonumber\\
&  \times\left(  \frac{kT_{e}}{\epsilon^{2}}\frac{\partial}{\partial\epsilon
}\epsilon^{4}\frac{\partial}{\partial\epsilon}+\frac{1}{\epsilon^{2}}%
\frac{\partial}{\partial\epsilon}\epsilon^{4}+2n\frac{\partial}{\partial
\epsilon}\epsilon^{2}\right)  n^{\prime}\Bigg\},\nonumber
\end{align}
where $T_{e}$\ is electron comoving temperature, $n_{e}$ is electron comoving
density, $m_e$ is electron mass, $\sigma_{\mathrm{T}}$ is Thompson cross section, $c$ is the speed of light, $n=n(\epsilon,\mu)$,
$n^{\prime}=n(\epsilon,\mu^{\prime})$, $do^{\prime}=d\mu^{\prime}d\phi
^{\prime}$, and $x=\sqrt{1-\mu^{2}}\sqrt{1-\mu^{\prime2}}\cos(\phi
-\phi^{\prime})+\mu\mu^{\prime}$. Here prime denotes angles of particles after scattering.

We integrate Eq. (\ref{Coll}) over $d\phi$. Then we introduce the Legendre
polynomials%
\begin{align}
n(\epsilon,\mu)  &  =\sum_{l}n_{l}(\epsilon)P_{l}(\mu),\\
P_{0}  &  =1,\quad P_{1}=\mu,\quad P_{2}=\frac{3\mu^{2}-1}{2},\quad
P_{3}=\frac{5\mu^{3}-3\mu}{2},\nonumber
\end{align}
and transform to a new variable $\rho\equiv n\epsilon^{2}$. Hence we arrive to
the final equation%

\begin{gather}
\frac{\partial\rho}{\partial r}+\frac{2\rho}{r}+\frac{1}{r(\beta+\mu)}%
\frac{\partial}{\partial\mu}\left[  (1-\mu^{2})(1+\beta\mu)\rho\right]
\label{Komp}\\
-\beta\frac{1-\mu^2}{r(\beta+\mu)}\frac{\partial(\epsilon\rho)}{\partial\epsilon}%
=\frac{\epsilon^{2}(\eta-\chi n)}{\Gamma(\beta+\mu)}\nonumber
\end{gather}
with%
\begin{gather}
\epsilon^{2}(\eta-\chi n)=n_{e}\sigma_{\mathrm{T}}\epsilon\Bigg\{\left(1-
\frac{2\epsilon}{m_{e}c^{2}}\right) \label{Coll2}\\
\times\left[  \rho_{0}+\frac{1}{10}\rho_{2}P_{2}-\rho\right]
\nonumber\\
+\frac{2\epsilon}{m_{e}c^{2}}\left(  \frac{2}{5}\rho_{1}P_{1}+\frac{3}{70}%
\rho_{3}P_{3}\right) \nonumber\\
-\frac{2kT_{e}}{m_{e}c^{2}}\left(  \frac{1}{5}\rho_{1}P_{1}\frac{3}{10}%
\rho_{2}P_{2}+-\frac{3}{35}\rho_{3}P_{3}\right) \nonumber\\
+\frac{1}{m_{e}c^{2}}\left[  \frac{\partial}{\partial\epsilon}\left(
kT_{e}\epsilon^{2}\frac{\partial}{\partial\epsilon}+\epsilon(\epsilon
-kT_{e})\right)  +2\rho\frac{\partial}{\partial\epsilon}\right]
\times\nonumber\\
\times\left(  \rho_{0}-\frac{2}{5}\rho_{1}P_{1}+\frac{1}{10}\rho_{2}%
P_{2}-\frac{3}{70}\rho_{3}P_{3}\right)  \Bigg\}.\nonumber
\end{gather}

In isotropic case $\rho=\rho_0$ and the integration of equation (\ref{Komp}) over angles gives the Kompaneets equation
for the variable $\rho_0$ as%
\begin{gather}
\frac{\partial\rho}{\partial r}+\frac{2\rho}{r}=\frac{\epsilon^{2}n_{e}%
\sigma_{\mathrm{T}}}{\Gamma m_{e}c^{2}}\left(  1-\frac{2\epsilon}{m_{e}c^{2}%
}\right) \\
\times\left[  \frac{\partial}{\partial\epsilon}\left(  kT_{e}\epsilon^{2}%
\frac{\partial}{\partial\epsilon}+\epsilon(\epsilon-kT_{e})\right)
+2\rho\frac{\partial}{\partial\epsilon}\right]  \rho.\nonumber
\end{gather}
As expected, the photon number conservation holds, that is $\int d\epsilon
r^{2}\rho=\mathrm{const}$. When angle dependence is taken into account photon
conservation gives%

\begin{equation}
\int d\mu d\epsilon(\beta+\mu)r^{2}\rho=\mathrm{const}. \label{conslaw}%
\end{equation}

Equation (\ref{Komp}) with collision integral (\ref{Coll2}) is integrated
numerically. We introduced the computational grid for angles and energy and applied lines method for evolutionary equations with coordinate $r$ playing the role of time. Implicit numerical scheme is used. Our scheme for the classical Kompaneets equation is similar to the one used in \cite{1997Ap.....40..227N}.
In numerical calculations one has to assume the temperature and number density
of electrons dependence on radius. Following \cite{2013ApJ...772...11R} we
adopt the comoving density profile
\begin{equation}
n_{e}=n_0 B \left(\frac{R_0}{R}\right)^2,
\end{equation}
where $B=\dot M c^2/L$ is the baryonic loading parameter, $L$ is luminosity at the base of the wind at radius $R_{0}$ and $\dot M$ is mass ejection rate.
For the comoving temperature we choose
\begin{equation}
T_{e}=T_0 B \left(\frac{R_0}{B R}\right)^k,\label{scalings}%
\end{equation}
where $T_0=(L/(16\pi \sigma_{SB}R_0^2))^{1/4}$, $\sigma_{SB}$ is Stefan-Boltzmann constant, $k$ is a constant.
This model with $k=2/3$ describes coasting steady relativistic wind.

The electron comoving density decreases with increasing laboratory radius, so
does the optical depth. Near the photosphere, where the optical depth reaches
unity, the coupling between photons and electrons weakens. We assume that
electron comoving temperature on the path of photons propagating outwards
is decreasing following eq. (\ref{scalings}), even when the outflow becomes optically thin. This effect is parametrized by the coefficient $k\geq 2/3$. Electrons are
described by the Maxwellian distribution function in the comoving frame with
the temperature given by the relation (\ref{scalings}).

Since initially the outflow is highly opaque, photons in the beginning have
thermal spectrum and isotropic distribution in the comoving reference frame.
Near the photosphere the coupling of photons to the medium is due to Compton
scattering on electrons.

Before proceeding with steady wind model, we tested our numerical scheme.
Considering a uniform optically thick medium with stimulated emission taken
into account and neglected, respectively, convergence to Planck and Wien
distributions of photons has been verified.

\section{Results}

\begin{figure}
\centering
\includegraphics[width=3.3in]{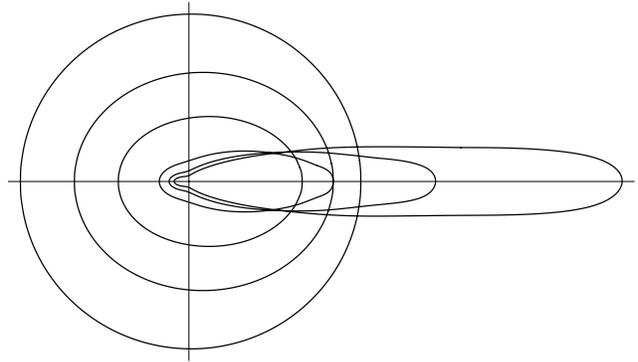} \caption{Anisotropy of photon
distribution in the comoving reference frame developing near the photosphere. When the outflow has high
optical depth the distribution is isotropic, but near the photosphere it
becomes increasingly anisotropic. The direction of medium motion is to the
right. The optical depth for these curves is $\tau=100, 10, 5, 1.4, 0.7, 0.3$ with increasing anisotropy.}%
\label{anisotropy}%
\end{figure}

The anisotropy of photon distribution near the photosphere in comoving
reference frame found by \cite{2011ApJ...737...68B} is an interesting effect
which we have also found in our approach. This result is illustrated in Fig.
\ref{anisotropy}.

This effect can be explained from the geometric point of view.\ Since
collisions tend to isotropize photons, only those photons which are undergoing
scatterings have nearly isotropic distribution. In contrast photons that
already experienced their last scattering have increasingly anisotropic
distribution in the comoving frame due to relativistic aberration. Hence the
local photon field at the photosphere which contains all photons, those which
continue to scatter and those which already propagate freely, becomes more and
more anisotropic.

Our computation is performed in the comoving reference frame. In order to find
the photon spectrum which a distant observer will detect we have to transform
the spectrum into the laboratory reference frame. One has to use the following transformations%

\begin{align}
N(\epsilon_{L})=2\pi\int_{0}^{1}d\mu_{L}r^{2}\mu_{L}n(\epsilon,\mu
))\epsilon_{L}^{2},\\
\epsilon=\Gamma(1-\beta\mu_{L})\epsilon_{L}, \quad\quad\mu=\frac{\mu_{L}%
-\beta}{1-\beta\mu_{L}},\label{transfs}%
\end{align}
where the subscript "$L$" denotes quantities in the laboratory frame.

Two asymptotic cases were found in \cite{2013ApJ...772...11R} to have very
different spectra from the photosphere:\ photon thick and photon thin ones.
Since the photon thin case is characterized by nearly Planck instantaneous
observed spectrum, here we concentrate on the photon thick case.

Consider steady relativistic wind with the following parameters:
\begin{equation}
L=10^{54}\text{ erg/s, }\Gamma=500\text{, }R_{0}=10^{8}\text{ cm}\text{, } k=2/3.%
\end{equation}
Given these values we find for initial optical depth%
\begin{equation}
\tau_{0}=\frac{\sigma_{\mathrm{T}}L}{4\pi m_{p}c^{3}R_{0}\Gamma}%
=2.2\times10^{10},
\end{equation}
where $\sigma_{\mathrm{T}}$\ is the Thomson cross section, and initial
temperature $T_{0}=1.2$ MeV. The photospheric radius is%
\begin{equation}
R_{ph}=\frac{\tau_{0}}{2\Gamma^{2}}R_{0}=4.4\times10^{12}\text{ cm.}%
\end{equation}

We start the computation at the beginning of the coasting phase with
$R_{i}=5\times10^{10}$ cm where the optical depth is $\tau_{i}\sim 10^{2}$ and
compute the spectrum until it settles to a static solution in the laboratory
reference frame well above the photospheric radius $R\gg R_{ph}$.

\begin{figure}
\centering
\includegraphics[width=3in]{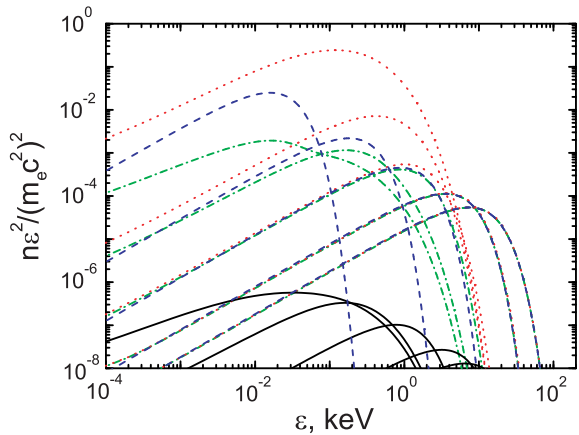} \caption{The spectrum of photons
($F_{\nu}$) in the comoving reference frame is shown for selected values of
radii: $5\times 10^{10}$, $10^{11}$, $10^{12}$, $10^{13}$ and $5\times10^{14}$ cm. Solid curves
show integrated spectrum over angles, dashed curves show the Planck spectrum with the temperature
following the relation (\ref{scalings}) with $k=2/3$ for comparison. Dotted curves show the spectra for $\mu=1$ (photons propagating forward in the comoving frame) and dash-dotted curves correspond to $\mu=-1$ (photons propagating backward in the comoving frame).}%
\label{comsp}%
\end{figure}
\begin{figure}
\centering
\includegraphics[width=3in]{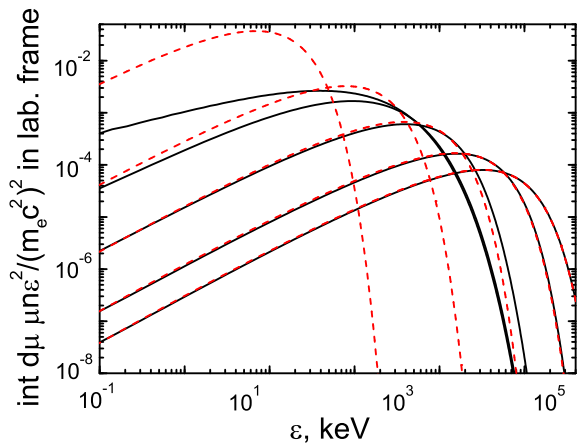} \caption{The spectrum of photons
($F_{\nu}$) transformed to the laboratory reference frame is shown for
selected values of radii, corresponding to Fig. \ref{comsp}. Solid curves show
the observed spectrum of photons integrated over angles, dashed curves show
the spectrum of photons arriving to a given radius with thermal spectrum in
the comoving reference frame, transformed to the laboratory reference frame.}%
\label{labsp}%
\end{figure}

In Fig. \ref{comsp} we show the evolution of photon spectrum in comoving
reference frame for selected values of radii starting at large optical depth
through the photosphere ending up at small optical depth $\tau\sim10^{-2}$.

While the spectrum in comoving frame initially keeps the Planck shape, it becomes distorted with decreasing
optical depth. It is also the consequence of interaction with
electrons which on average have smaller energy than photons do, see Fig. \ref{comsp}. In fact, near
the photosphere the average energy of photons saturates
\cite{2011ApJ...737...68B}, but the one of electrons continues to decrease.
When this spectrum is transformed to the laboratory frame using Eq. (\ref{transfs}) we find increasing deviations from Planck spectrum essentially due to contribution from different angles. Clearly this effect becomes less and less prominent with decreasing optical depth. The final shape of the spectrum is reached for $\tau>10^{-2}$. Our results for the steady coasting relativistic wind agree quantitatively with those obtained by \cite{2013ApJ...767..139B} and by \cite{2013ApJ...772...11R}.

In Fig. \ref{comsp} we also show the spectra of photons propagating forward (for which the angle $\theta$ between the momentum of the photon and the radius vector vanishes, dotted curve) and those propagating backward (for which $\theta=\pi$, dash-dotted curve). While photons propagating forward dominate in the spectrum due to Doppler effect, the contribution of photons with $\theta>0$ changes the shape of the spectrum compared to the black body one, especially near its peak.

We also computed observed spectrum with different values of $k$ parameter. The values of $k>2/3$ imply that the temperature in the wind decreases with radius faster than the one given by the steady solution. More rapid change of temperature with radius may mimic some more realistic profiles of finite radial extension of the outflow, see e.g. \cite{1993MNRAS.263..861P}. Photons propagating in such outflows scatter on electrons with lower temperature, which leads to additional comptonization of the spectrum. At large optical depths photon spectrum adjusts to the electron temperature. With decreasing optical depth only low energy part of the spectrum is affected by comptonization.
\begin{figure}
\centering
\includegraphics[width=3.0in]{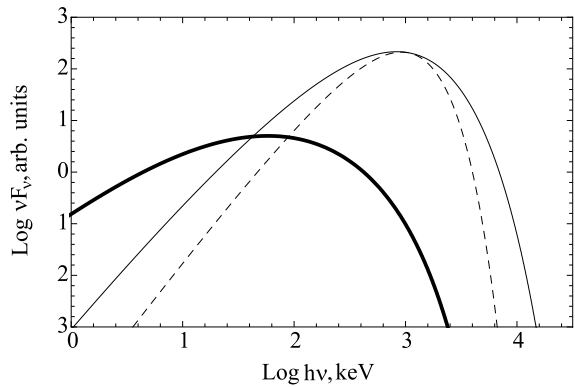} \caption{The final observed spectrum of photons
($\nu F_{\nu}$) is shown for two values of $k$ parameter: $k=2/3$ (thin curve) and $k=2$ (thick curve). For comparison Planck spectrum is shown by dashed curve.}%
\label{spectra}%
\end{figure}
In Fig. \ref{spectra} we show for comparison two final observed spectra $\nu F(\nu)\propto n\epsilon^4$ with different values of $k$, namely $k=2/3$ and $k=2$. It is interesting to note that the number of photons in the low energy part of the spectrum increases with increasing $k$. This low energy part of the spectrum can be fit with a power law. Performing the fit in the energy range between 1 and 10 keV we found that the power law index changes as follows
\begin{align}
	\alpha\simeq 1-k.
\end{align}
This power law index $\alpha$ actually coincides with the one used in the Band phenomenological spectrum \cite{1993ApJ...413..281B}. Thus we find that Compton scattering near the photosphere can change the low energy part of observed spectrum. In particular, the photon index $\alpha$ is shown to depend on the index $k$ which parametrizes the electron temperature dependence on radius.

In our calculations the stationary wind model is used. The observed durations of GRBs range from a fraction of second to thousands of seconds, hence the finite duration effects have to be taken into account for theoretical calculations of light curves and spectra of the photospheric emission. Such finite duration effects give rise to a new classification of relativistic outflows, which is discussed in details by \cite{2013ApJ...772...11R}, see also \cite{2013ApJ...767..139B}.

Besides one has to keep in mind that the prompt emission of GRBs contains not only photospheric component but in most cases also non-thermal component originating from optically thin regions of the outflow. However, unambiguous identification of the photospheric component in GRBs is crucial since it provides basic information about the physical parameters of GRBs, see e.g. \cite{Iyyani21082013}.

\section{Conclusions}

We considered relativistic steady wind and computed the observed spectrum of
its photospheric emission. For this goal we solved the equation of radiative
transfer in comoving reference frame with exact Compton collision term in
Fokker-Planck approximation, including the effect of anisotropy of the photon
field. We obtained the photon spectra in the laboratory frame, as seen by a
distant observer.

We confirmed the result of \cite{2011ApJ...737...68B} indicating the presence of
strong anisotropy developed in the comoving frame at the photosphere. Some
qualitative differences in our results are due to the fact that he used
approximate collision term, instead of our exact one.

The observed spectrum from the photosphere of the steady coasting relativistic wind is found in agreement with previous studies. We also found that when the temperature decreases with radius faster than in the case of a steady wind the low energy part of the observed spectrum changes significantly with respect to the Planck function. In particular, for the temperature profile $T\propto r^{-2}$ we found for the Band index $\alpha\simeq-1$, in agreement with typical low energy photon index observed in GRBs.

\bibliographystyle{mn2e}

\end{document}